\documentclass[fleqn,usenatbib]{mnras}

\usepackage{newtxtext,newtxmath}

\usepackage[T1]{fontenc}

\DeclareRobustCommand{\VAN}[3]{#2}
\let\VANthebibliography\thebibliography
\def\thebibliography{\DeclareRobustCommand{\VAN}[3]{##3}\VANthebibliography}

\usepackage{graphicx}	
\usepackage{amsmath}	
\usepackage{color}
\usepackage{float}
\usepackage{ulem}
\usepackage{orcidlink}
\usepackage{listings}

\newcommand{\roger}{\textsc{roger}}
\newcommand{\rogerii}{\textsc{roger\ v2.0}}

\newcommand{\pyroger}{\textsc{pyroger}}
\newcommand{\hmsun}{h^{-1} M_\odot}
\newcommand{\hmpc}{h^{-1}{\rm Mpc}}

\definecolor{codegreen}{rgb}{0,0.6,0}
\definecolor{codegray}{rgb}{0.5,0.5,0.5}
\definecolor{codepurple}{rgb}{0.58,0,0.82}
\definecolor{backcolour}{rgb}{0.95,0.95,0.92}

\lstdefinestyle{mystyle}{
  backgroundcolor=\color{backcolour},   commentstyle=\color{codegreen},
  keywordstyle=\color{magenta},
  numberstyle=\tiny\color{codegray},
  stringstyle=\color{codepurple},
  basicstyle=\ttfamily\footnotesize,
  breakatwhitespace=false,         
  breaklines=true,                 
  captionpos=b,                    
  keepspaces=true,                 
  numbers=left,                    
  numbersep=5pt,                  
  showspaces=false,                
  showstringspaces=false,
  showtabs=false,                  
  tabsize=2
}
\title[\rogerii{}]{Reconstructing orbits of galaxies in extreme regions (\rogerii{}): an extension to intermediate mass systems}

\author[M. de los Rios et al.]{
Martín de los Rios$^{1}$\thanks{E-mail: martindelosrios13@gmail.com}\orcidlink{0000-0003-2190-2196},
Héctor J. Martínez$^{1,2}$\orcidlink{0000-0003-0477-5412},
Andrés N. Ruiz$^{1,2}$\orcidlink{0000-0001-5035-4913},
Selene Levis$^{1,3}$\orcidlink{0000-0003-1887-776X},
Valeria Coenda$^{1,2}$\orcidlink{0000-0001-5262-3822}\newauthor  
and Hernán Muriel$^{1,2}$\orcidlink{0000-0002-7305-9500}
\\
$^{1}$Instituto de Astronomía Teórica y Experimental, CONICET-UNC, Laprida 854, X5000BGR, Córdoba, Argentina\\
$^{2}$Observatorio Astronómico, Universidad Nacional de Córdoba, Laprida 854, X5000BGR, Córdoba, Argentina\\
$^{3}$Facultad de Matemática, Astronomía, Física y Computación, Universidad Nacional de Córdoba, Av. Medina Allende s/n, X5000HUA, Córdoba, Argentina
}
\date{Accepted XXX. Received YYY; in original form ZZZ}

\pubyear{\the\year{}}

\begin{document}
\label{firstpage}
\pagerange{\pageref{firstpage}--\pageref{lastpage}}
\maketitle

\begin{abstract}
In this paper, we present an updated version of the \roger{} code, called \rogerii{}, developed to 
perform the orbital classification of galaxies residing in and around galaxy groups 
and clusters. In addition to the projected phase-space coordinates, the new version 
incorporates the host halo mass as an additional input parameter. Although the 
inclusion of the halo mass leads to only modest changes in the classification, it 
contributes to improving the overall robustness of the method. We also extend the 
range of host halo masses over which \roger{} can be applied, enabling the analysis 
of systems with masses down to $10^{13.5}h^{-1}M_{\odot}$.
We further provide a Python implementation of the new code, which will be made 
publicly available. This implementation enables users to efficiently and robustly 
classify arbitrary galaxy samples using either version of the method. Moreover, it 
allows users to train a customized classifier on an alternative training set, 
providing the flexibility to adapt the method to different datasets and scientific 
applications.
\end{abstract}


\section{Introduction}
\label{sec:intro}

The Projected Phase Space Diagram (PPSD) provides a valuable diagnostic
framework to characterise the kinematic behaviour and dynamical state of galaxies
in systems such as clusters and groups. The PPSD is defined by two coordinates:
the projected cluster-centric (or group-centric) distance ($R_\mathrm{p}$) 
in units of a characteristic size of the system, such as the 
radius that encloses an overdensity greater than 200 times the 
critical density of the Universe ($R_{200}$), and the line-of-sight velocity 
($\Delta V$) in units of the velocity dispersion of
the system ($\sigma$). This diagram has been widely used to analyse various 
aspects of galaxy evolution. Early works investigate how the location in the PPSD
relates to the decline of star formation in backsplash galaxies \citep{Mahajan:2011} 
and the quenching of galaxies at $z\sim 1$ \citep{Muzzin:2014}. Subsequent studies examine the
properties of high- and low-velocity galaxies in cluster outskirts
\citep{Muriel:2014}, and characterise the effects of ram pressure and gas
fraction in clusters \citep{HF:2014, Jaffe:2015}. More recently, the diagram
provides the basis for analyses exploring the connection between star formation
rate and orbital history \citep{Oman2013,Oman:2016}, stellar mass growth as a function of
infall time \citep{Smith:2019}, and quenching and morphology in groups or
clusters \citep{Martinez:2023, Oxland:2024, Muriel:2025}. Concurrently, other
works focus on the environmental influence on the properties of barred galaxies
 \citep{Aguerri:2023}, the evolution of colour fractions \citep{Sampaio:2024},
 the impact on black hole growth \citep{Munoz:2024}, and the classification of
 merging galaxies \citep{Kim:2024}.

There are several methods to classify galaxies in the PPSD. For instance,
\citet{Rhee:2017} use cosmological hydrodynamical N-body simulations of groups
and clusters covering the virial mass range $M_\mathrm{vir}=5.3\times10^{13}-9.2\times 10^{14} M_{\odot}$, 
to define specific regions of the PPSD, where galaxies with different first infall
times preferentially reside. Using the same cosmological simulations as
\citet{Rhee:2017}, \citet{Pasquali:2019} define eight zones of constant mean
infall time to assess environmental influences on satellite galaxies. These
zones are delineated by analytical curves that characterise the timing of system
infall. 

\roger{} was developed to connect the observational properties of galaxies with its real dynamical orbits. The first version of the code was introduced in \cite{delosRios:2021}, originally trained for analyzing massive clusters ($M_{200}\ge 10^{15}\, h^{-1} M_{\odot}$, where $M_{200}$ is defined 
as the mass enclosed in a spherical volume of radius $R_{200}$), 
isolated galaxy clusters drawn 
from the \textsc{MultiDark Planck} 2 simulation \citep{prada_mdpl_2012,klypin_mdpl2_2016}. Specifically, 
\roger{} maps a galaxy’s two-dimensional (2D) PPSD position to its underlying
three-dimensional (3D) orbital class and calculates the
probability of a galaxy belonging to one of five categories: cluster members, backsplash galaxies, recent infallers, infaling, and
interloopers. In the second paper of the \roger{} series, \citet{Coenda:2022} (\roger\ II) 
studied how contamination from misclassifications biases the interpretation of galaxy properties. They emphasized that 
distinguishing
between red and blue populations is essential for improving observational
precision, concluding that 2D analyses yield reliable results only for recent infaller, infalling,
and interlooper populations within the blue galaxy sample.
To diminish the impact of the misclassifications, 
\citet{Martinez:2025} introduced a method which uses the confusion matrix to statistically correct the distribution of galaxy properties. This
approach is applicable to any PPSD classification and provides a robust way to
mitigate contamination. It is worth noting that, given its flexibility
and applicability to a broad range of PPSD classification problems,
this work was not originally presented as part of the \roger\ series.
Nevertheless, its development represents a natural continuation of
the research presented in the previous papers of the series. 
Continuing the \roger{} series, the code has also been applied to observational data to investigate the environmental effects on galaxies. In particular, it was used in
\citet{Martinez:2023} (\roger\ III) to study morphological and star-formation quenching transitions in massive X-ray-selected SDSS clusters. This study found evidence that galaxies can experience both a suppression of star formation and morphological changes after a single passage through the cluster, with quenching occurring on shorter timescales than morphological transformation.
Furthermore, \roger{} was used to investigate the environmental evolution of galaxies in 35 OmegaWINGS clusters in \citet{Muriel:2025} (\roger{}\ IV). Their results suggest that both, morphological transformation, and the quenching of star formation, begin shortly after galaxies enter the cluster. However, even after $2$--$3$ Gyr, these processes are not yet complete, suggesting that the effects of the cluster environment continue to develop over several Gyrs after infall.

However, these previous studies were limited to high-mass regimes, as the method was
trained exclusively on massive clusters. Consequently, this work aims to extend
\roger{} to a broader mass range by using a training sample of clusters and
groups with masses $M_{200} \ge 10^{13.5}\hmsun$. In addition, we incorporate
the system's mass as a new input variable and study how it affects the PPSD and
the classification performance. This extension became very important in light of
the incoming surveys such as \textit{Euclid} \citep{euclid_2011,euclid_2025},
the \textit{Dark Energy Spectroscopic Instrument}
\citep[DESI,][]{desi_2016,desi_dr1_2026} and the \textit{Chilean Cluster Galaxy
Evolution Survey} \citep[CHANCES,][]{chances_2025}. A first application of this
extended version of \rogerii{} was implemented by \cite{levis2026} to study the impact of hot intra-group medium on galaxy evolution in
a sample of galaxy groups from the Galaxy And Mass Assembly survey
\citep[GAMA,][]{Driver:2009,Driver:2011}.

This paper is organised as follows: in Sect. \ref{sec:data} we present the
dataset of simulated galaxies we use to train and test this new version of
\rogerii{}, and also analyze how the different orbital classes populate the PPSD as
a function of halo mass; in Sect. \ref{sec:model} we train and test the model;
finally, in Sect. \ref{sec:conclu} we present our conclusions.

\section{Data}
\label{sec:data}

\subsection{The \textsc{MDPL2-SAG} galaxy catalog}

The main data sample used in this work comes from the galaxy catalog \textsc{MDPL2-SAG}
\footnote{\url{https://doi.org/10.17876/cosmosim/mdpl2/007}} \citep{Knebe2018}. This
catalog, which is publicly available in the
\textsc{CosmoSim}\footnote{\url{https://www.cosmosim.org}} database
\citep{riebe_multidark_2013}, was constructed by combining the \textsc{MultiDark
Planck 2}\footnote{\url{https://doi.org/10.17876/cosmosim/mdpl2}}
cosmological simulation \citep[\textsc{MDPL2},][]{prada_mdpl_2012,klypin_mdpl2_2016} and
the semi-analytic model of galaxy formation and evolution
\textsc{SAG} \citep{cora_sag_2018}. 

The simulation \textsc{MDPL2} has $3840^3$ dark matter particles in a periodic cubic
volume with a side length of $1000\hmpc$ and cosmological parameters consistent
with Planck \citep{planck_cosmology_2016}. The simulation counts with 127
snapshots between $z=17$ and $z=0$, and was evolved using an optimized version of the public code
\textsc{gadget2} \citep{Springel2005_gadget2}. Dark matter halos were identified
using the \textsc{Rockstar} halo finder \citep{behroozi_rockstar_2013} and
merger trees were constructed with \textsc{ConsistentTrees}
\citep{behroozi_trees_2013}. The halo catalogs and merger trees of \textsc{MDPL2} are also
available at \textsc{CosmoSim}.

The dark matter halos were populated with semi-analytic galaxies using
\textsc{SAG}. This model includes all the relevant physical processes in galaxy
formation: radiative cooling of gas, star formation, feedback by supernovae
explosions, follow-up of chemical enrichment of gas, supermassive black holes
and their respective feedback, galaxy mergers, ram pressure and tidal stripping.
For a complete description of the implementation of these processes, we refer the 
reader to \citet{cora_sag_2018,Cora2019_sag}.

\subsection{Selection of galaxy systems and galaxies}

The halo catalogue of \textsc{MDPL2} simulation comprises $\sim 1.27 \times 10^7$
halos with more than 20 particles at $z=0$. From these halos, we select those with
a mass $M_{\rm 200} \ge 10^{13.5} \hmsun$.
Additionally, we impose an isolation criterion, where no halo can have 
a neighbour more massive than $0.1\times M_{\rm 200}$ within $5\times R_{\rm 200}$.
This isolation imposition is to avoid serious perturbations in galaxy orbits in the
vicinity of halos due to mergers or heavy interactions. 

For each of these clusters, we select all galaxies at $z=0$ whose
comoving position and peculiar velocity relatives to the cluster, $\Delta
\mathbf{r}$ and $\Delta \mathbf{v}$,
satisfy, for at least one of the three cartesian axes ($i=x,y,z$) in the simulation 
box, that the projected distance is
\begin{equation}
   R_{\mathrm{p}, i}\equiv |\Delta \mathbf{r}-(\Delta \mathbf{r}\cdot \hat{\imath})\, \hat{\imath}| \le 3\times R_{200},
\end{equation}
and the line-of-sight velocity is
\begin{equation}
  |\Delta V_i|\equiv | (\Delta \mathbf{v} + H_0\Delta\mathbf{r})\cdot \hat{\imath}
  | \le 3\times \sigma_i,
\end{equation}
where $\hat{\imath}$ is the versor in the $i-$coordinate, and $\sigma_i$ is the 
one dimensional velocity dispersion of the cluster along that direction,
and $H_0 = 100h$ km s$^{-1}$ Mpc$^{-1}$ is the Hubble parameter today\footnote{Galaxies and clusters
positions are in $\hmpc$ units, so is not necessary to specify the value of
$h$.}. The values of $\sigma_i$ were calculated with all galaxies inside
$R_{200}$ using the biweight estimator presented in \citet{Beers90}.

Once the galaxies are selected at $z=0$, we track them 
up to $z \sim 2$ (from snapshot 125 to 75) using the \texttt{galaxystaticid} provided in the data base, a galaxy
identification that remains unchanged across all snapshots. Besides the
galaxies located in the surroundings of the clusters, we include
interlopers, i.e. galaxies unrelated to clusters but that will appear in or
around them when we look at their positions in projection. These interlopers
constitute an important source of contamination in 
the observed projected phase-space diagram, 
see Sect. \ref{sec:orbits} for the details
of how this sample is constructed. 

To include only reliable galaxies both in properties and
orbits, we consider galaxies with total stellar masses $M_\star \ge 10^9 \hmsun$
\citep{Knebe2018} and we do not take into account orphan galaxies, i.e.
satellite galaxies that lost their dark matter host halo due to dynamical
friction and/or resolution effects \citep{cora_sag_2018}. Our final
\textsc{MDPL2-SAG} sample comprises $36,840$ clusters/groups 
and $6,297,699$ galaxies.

\subsection{Orbit classification}
\label{sec:orbits}

Following \citet{delosRios:2021}, we classify all selected
galaxies into five types according to their orbits around clusters. These types
are defined as follows:

\begin{itemize}

    \item \textbf{Cluster members} (CL): galaxies that are now orbiting the
    cluster and have crossed $R_{200}$ several times in their lifetime. Most of
    them are found inside $R_{200}$. 
    
    \item \textbf{Recent infallers} (RI): galaxies that have crossed $R_{200}$
    only once, and in their way in, in the last 2 Gyr. 
    Some of these galaxies may go beyond $R_{200}$ in the future. 
    
    \item  \textbf{Backsplash galaxies} (BS): galaxies that are located beyond
    $R_{200}$ at $z=0$ but have passed through the cluster once in the past,
    crossing $R_{200}$ exactly twice on their way in and out. According to this
    definition, some recent infallers will become backsplash and some backsplash will become cluster members in the
    future. 
    
    \item \textbf{Infalling galaxies} (IN): galaxies that have been outside
    $R_{200}$ during their entire lifetime but are now falling toward the
    cluster, as shown by their negative radial velocities relative to the
    cluster. 
    
    \item \textbf{Interlopers} (IL): galaxies located beyond $R_{200}$ and with
    positive radial velocities with respect to the center of the cluster. In
    contrast to infalling galaxies, these objects will not fall into the cluster in the
    future and are not physically associated to it. We keep them to emulate 
    the contamination in the observed PPSD. 
    
\end{itemize}

\subsection{How galaxy classes populate the PPSD as a function of halo mass}
\label{sec:ppsd}

The present improvement upon the original \roger\ adds the mass of the parent
system as a third parameter alongside the position of a galaxy in the PPSD.  It
is worth addressing the issue of whether the distribution of the classes $1-5$
over the PPSD depends on halo mass. 

\begin{figure}
    \centering
    \includegraphics[width=1.0\linewidth]{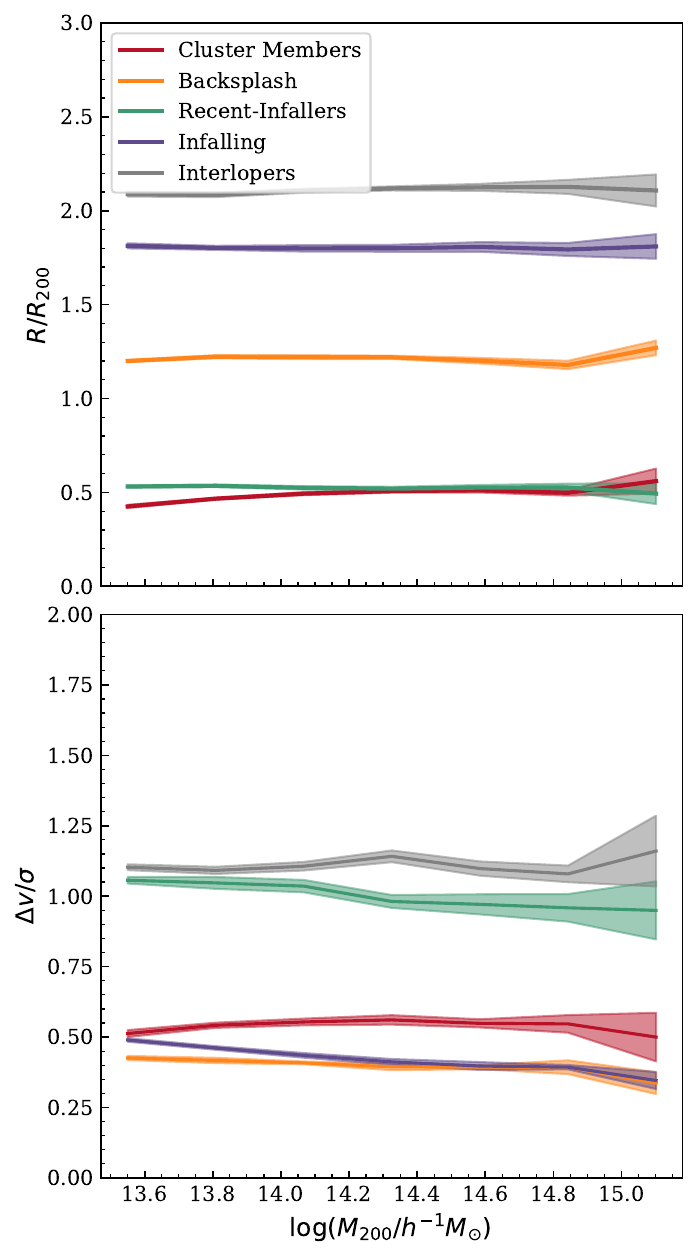}
    \caption{The halo mass dependence of the median values of the PPSD coordinates
    for the five classes. Upper panel shows the median of the projected 
    distance in units of $R_{200}$ as a function of $M_{200}$. Lower panel
    shows the relative line of sight velocity in units of $\sigma$. Error
bars correspond to a standard deviation computed with a bootstrap resampling
method.}
    \label{fig:medians}
\end{figure}

For each class, we show in Fig. \ref{fig:medians} the mass dependence of the median
values of $R_\mathrm{p}/R_{200}$ (upper panel) and $|\Delta V|/\sigma$ (lower
panel), respectively. We include in our analyses only those galaxies
with  $R_\mathrm{p}\leq 3\times R_{200}$ in projection and with velocities 
relative to the haloes $|\Delta V|\leq 3\times \sigma$. Each galaxy can, in 
principle, contribute to Fig \ref{fig:medians} up to three times, the line of 
sight direction being alternatively the $x$, $y$, and $z$ axes. 
For all cases, the median values of the PPSD coordinates do not depend on mass, or
the mass dependence is mild.
Regarding the median of the $R_\mathrm{p}/R_{200}$ coordinate, cluster galaxies show a
slight growing tendency with mass. On the other hand, the only tendencies of 
the median of $|\Delta V|/\sigma$ with mass are seen for recent infallers and infalling galaxies; in
both cases, a subtle decline with mass is observed.
All this suggests that halo mass plays, at best, a secondary role on the
way the different classes spread over the PPSD.

\begin{figure}
    \centering
    \includegraphics[width=1.0\linewidth]{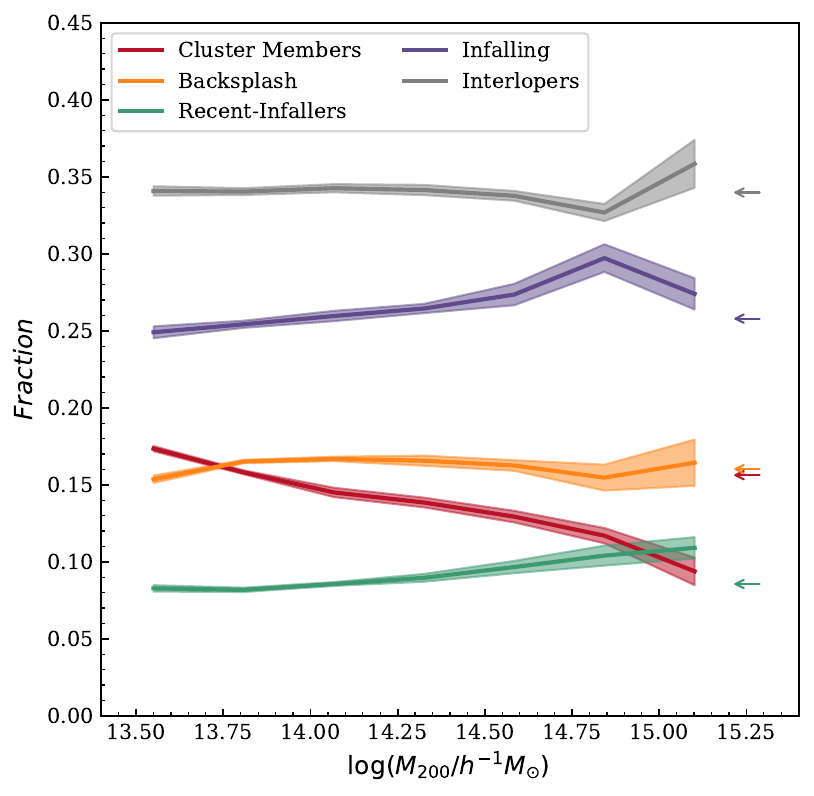}
    \caption{Fraction of galaxies in each of the five classes as a function
    of the halo mass. Arrows correspond to the respective global median values.
    Error bars correspond to a standard deviation computed with a bootstrap resampling method.}
    \label{fig:fractions}
\end{figure}

We show in Fig. \ref{fig:fractions} the fraction of galaxies in each class
as a function of halo mass. Galaxies contributing to this plot are the same as in
Fig. \ref{fig:medians}. The fraction of cluster galaxies decreases with mass, from
$\sim 0.17$ at $10^{13.5}\hmsun$ to $\sim 0.1$ at $10^{15}\hmsun$, 
Opposite trends are seen for recent infallers and infalling galaxies, in both cases the fraction increases 
with mass: from $\sim 0.08$ to $\sim 0.11$ in the case of recent infallers, and from $\sim 0.25$ to $\sim 0.28$ for infalling galaxies.
The remaining two classes do not exhibit a significant variation with mass. These trends 
can be explained in terms of massive systems being found in denser environments and also, 
they are still accreting galaxies and smaller systems of galaxies at a higher rate than lesser
massive systems.

\begin{figure*}
    \centering
    \includegraphics[width=\linewidth]{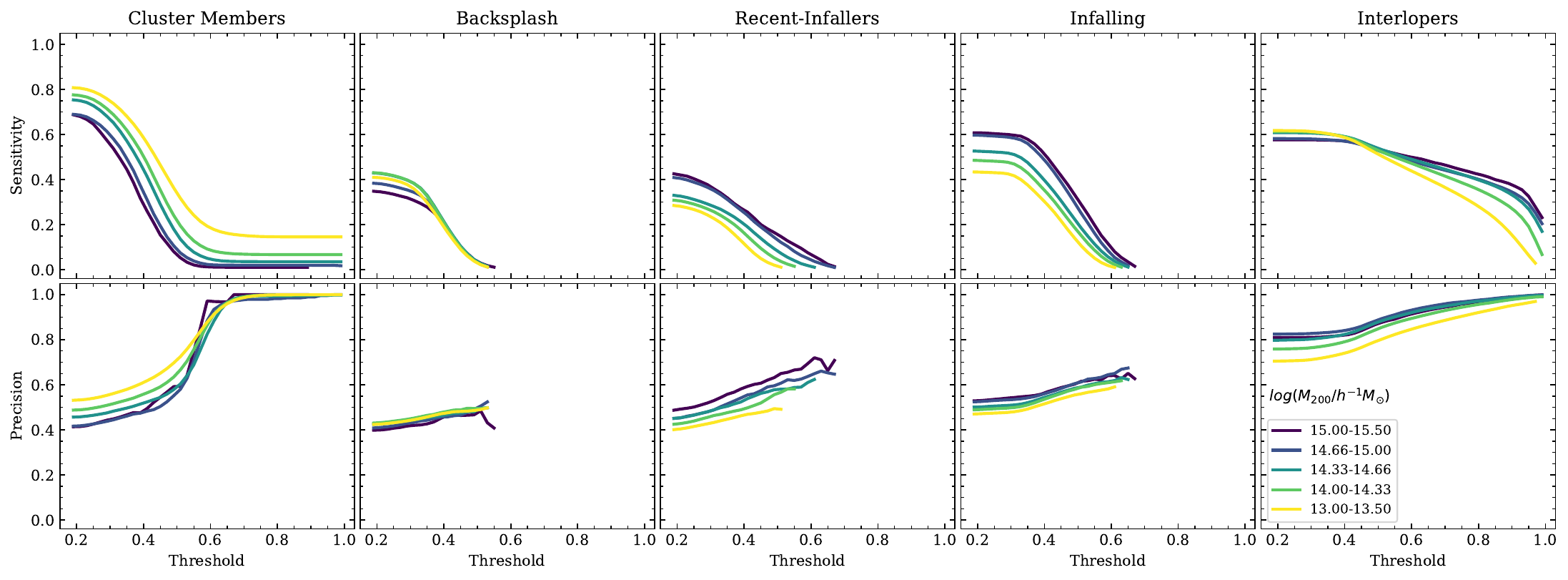}
    \caption{
    Top panels: Sensitivity as a function of threshold. The color of each curve corresponds to a specific 
    halo mass bin, as indicated in the fifth panel (interloopers). Bottom panels: Precision as a function of 
    threshold. Although, in principle, the threshold can span the range $0.2$–$1$ according to our 
    classification scheme, this interval is not always fully explored in practice, either because of 
    limited number statistics or because certain threshold values are not reached by the computed 
    probabilities.}
    \label{fig:sp_t}
\end{figure*}

\begin{figure*}
    \centering
    \includegraphics[width=\linewidth]{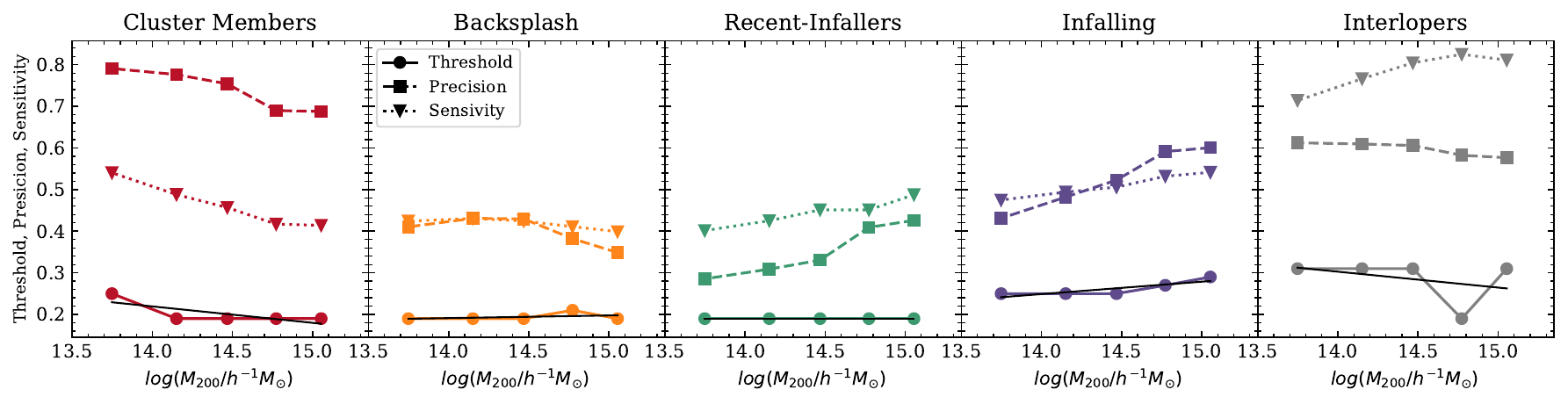}
    \caption{The threshold (solid lines) that maximises simultaneously sensitivity and precision as a function of halo mass.
    The associated sensitivity and precision are shown as dotted and dashed lines, respectively. 
    }
    \label{fig:tps}
\end{figure*}

\begin{figure*}
    \centering
    \includegraphics[width=0.98\linewidth]{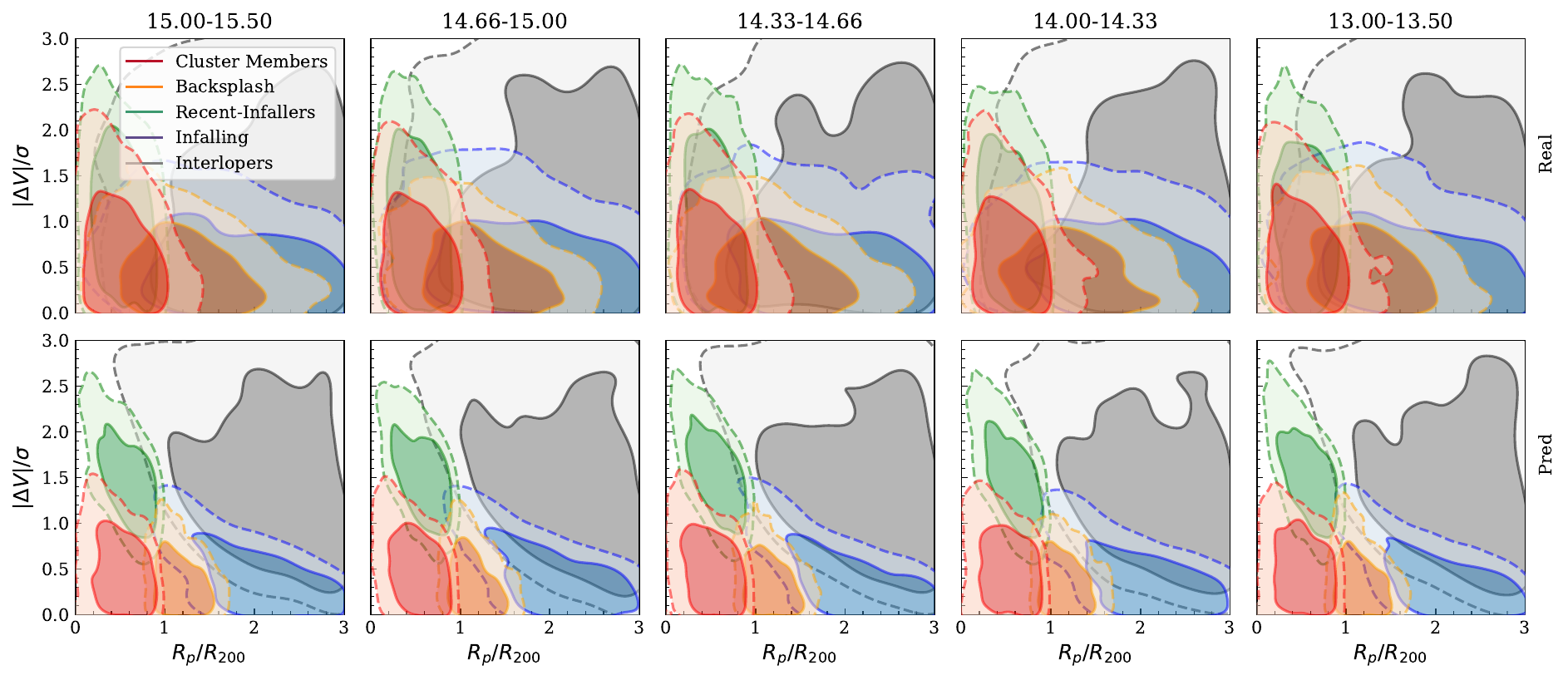}
    \caption{PPSD regions occupied by the different real classes (upper panels) and predicted classes (lower panels). Solid and dashed lines correspond to 1$\sigma$ and 2$\sigma$ regions respectively.}
    \label{fig:distros}
\end{figure*}

\begin{figure}
    \centering
    \includegraphics[width=1\linewidth]{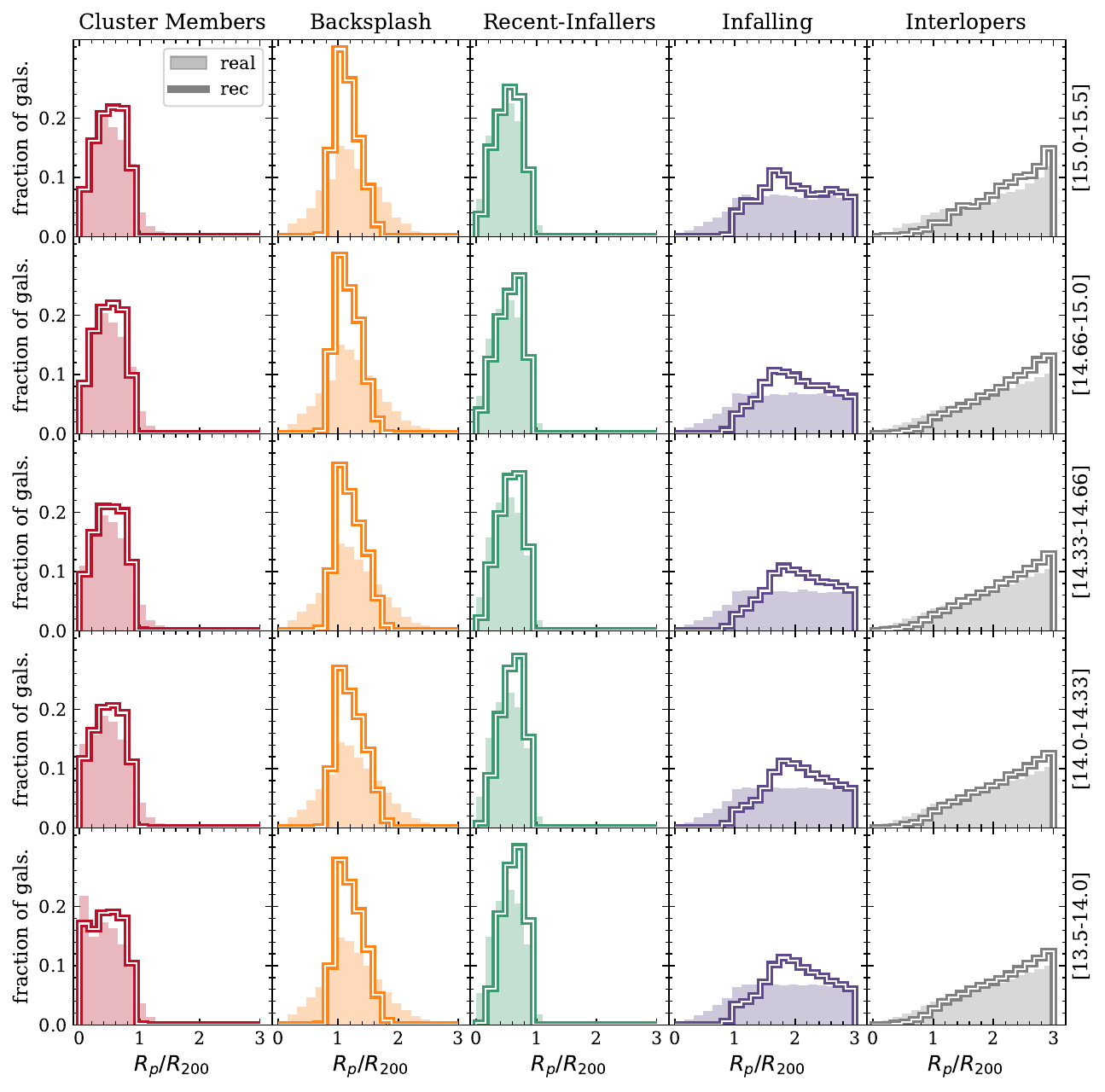}
    \caption{Normalised distributions of the projected distance in units of $R_{200}$. Shaded histograms are the real distributions, coloured-white lines are the predicted distributions using the mass dependent thresholds of Sect. \ref{sec:test}. Columns are the classes as quoted on top, rows correspond to the halo mass bins as quoted on the right.}
    \label{fig:hist_rp}
\end{figure}

\begin{figure}
    \centering
    \includegraphics[width=1\linewidth]{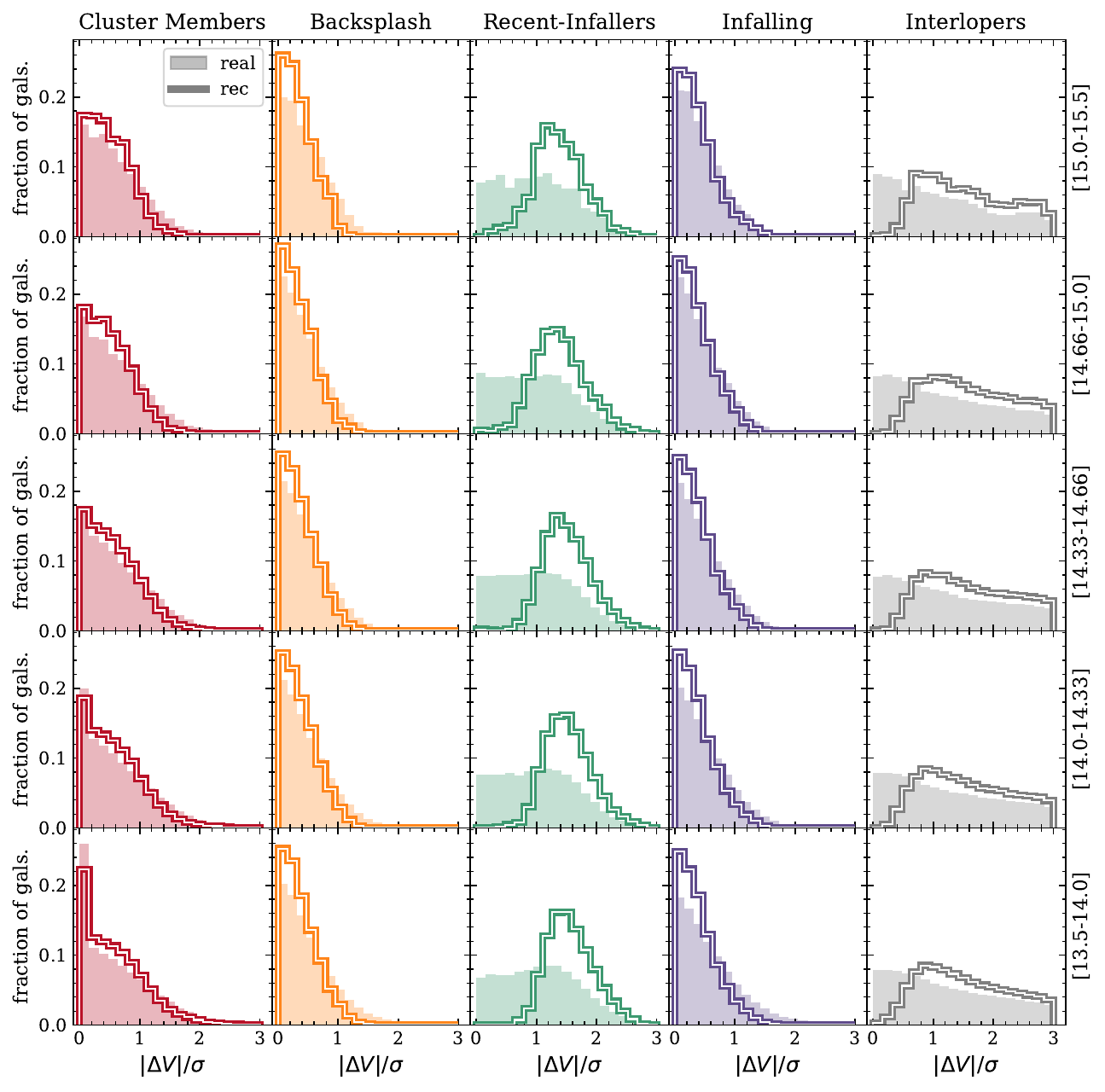}
    \caption{Normalised distributions of the velocity relative to the systems in units of $\sigma$. Shaded histograms are the real distributions, coloured-white lines are the predicted distributions using the mass dependent thresholds of Sect. \ref{sec:test}. Columns are the classes as quoted on top, rows correspond to the halo mass bins as quoted on the right.}
    \label{fig:hist_dv}
\end{figure}

\begin{figure*}
    \centering
    \includegraphics[width=\linewidth]{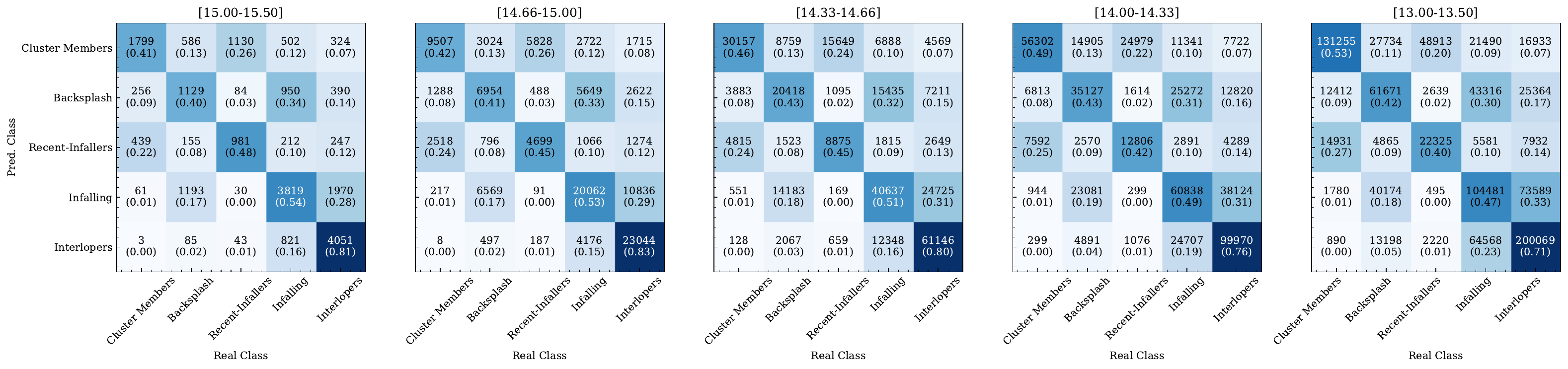}
    \caption{Confusion matrices resulting of the use of the halo mass dependent threshold of Sect. \ref{sec:test}. Columns are actual classes while rows are predicted classes. 
    Rows are normalized to illustrate the fraction of each actual class (inset numbers) that contributes to a specific predicted class. The halo mass range is quoted above each panel.}
    \label{fig:conf}
\end{figure*}

\section{The model}
\label{sec:model}

In this work, we present \rogerii{}\footnote{\url{https://github.com/Martindelosrios/pyROGER}}, an updated version of the dynamical classification framework introduced in \cite{delosRios:2021}. Building upon the original methodology, \rogerii{} incorporates the mass of the host cluster as an additional feature in the classification of individual galaxies.
We train two well-established and widely used 
techniques, namely K-Nearest Neighbors, and Random Forests \citep{rf}. For a thorough description of 
these methods we refer the reader to \cite{murphy2012machine}.

Following standard practice in supervised learning, the original dataset 
is first partitioned into two mutually exclusive subsets: a training set (comprised of the $80\%$ of the total galaxy sample) 
that will be used to train the models, and a testing set (comprised of the remaining $20\%$ of the galaxy sample), reserved for 
assessing the final performance after the training phase. 
Each cluster and galaxy in our samples contributes three times to  
either, the training or the testing subsets, as we project the three 
dimensional positions and velocities into the PPSD using three different
line-of-sight projections corresponding to each of the 
three independent axes of the simulation box: $x$, $y$, and $z$.

\subsection{Training the model}
\label{sec:training}

The training procedure involves the following input information from each 
galaxy in our training subset: 
\begin{enumerate}
    \item parent halo's $M_{200}$;
    \item galaxy's real class: cluster members, recent infallers, backsplash, infalling, or interloopers;
    \item galaxy's PPSD position: $(R_\mathrm{p}/R_{200}, |\Delta 
    V|/\sigma)$. 
\end{enumerate}
The training process 'teaches' the model how to compute the probability
of a galaxy of being of each of the five classes out of the three 
input parametres: the parent halo's mass, and the galaxy's position in the PPSD.

As the training procedure is intrinsically stochastic, we trained $10$ instances of each machine learning model.
For each model we perform a bootstrapping of the training set by randomly selecting observations from the original training set.  At testing time, for each galaxy we have $10$ estimated probabilities (one for each boostrapped instance), from which we compute the average probability of belonging to each class.

All the models were trained using the \textsc{sklearn} \citep{scikit-learn} 
python library, a well-known machine learning library and are available in the public github repository 
\footnote{\url{https://github.com/Martindelosrios/pyROGER}}.

In section \ref{app:pyroger} we include a brief tutorial on how to 
install and use our models.

\subsection{Testing the model}
\label{sec:test}

As in \citet{delosRios:2021}, our preferred method is KNN due to its simplicity. 
The other two methods provide very similar results and are not discussed further 
here. Nevertheless, both, SVM and Random Forest methods are included in the
\rogerii\ publicly available package and can be used if preferred by the user.

After \rogerii\ is trained, we use the control sample to conduct a number
of tests aimed to assess the performance of the method. We recall that \rogerii\
computes, out of a galaxy's position in the PPSD, the probabilities of it being of 
any of the 5 classes: $p_i$, $i=1,\ldots, 5$. There is no unique way of classifying
galaxies according to these probabilities, therefore, users have to adopt a classification 
criterion that suits their purposes. Among the various possibilities, the user can 
utilize the 
$p_i$ as weights in weighted statistics, 
adopt the class corresponding to the 
highest $p_i$ value, setting thresholds in the $p_i$ values, etc.
Our particular choice is to classify a galaxy as belonging to the
$j-$class (hereafter predicted class) if two conditions 
are met: (i) the maximum value of its $p_i$ corresponds to $i=j$, and (ii) $p_j>T_j$,
where $T_j$ is a threshold that can take numerical values from $0.2$ to $1$. This 
approach has been used before in \citet{Coenda:2022} and 
\citet{Martinez:2023,Martinez:2025}.

We use two standard measures of how good a classification scheme works
in terms of true/false positive/negative classifications, namely, precision 
and sensitivity. Precision is defined as the
ratio of true positive classifications over the sum of true positive and false 
positive classifications. Sensitivity is the ratio of true positive classifications
over the sum of true positive and false negative classifications. 
We show in Fig. \ref{fig:sp_t} sensitivity and precision as a function of threshold
for each of the 5 classes and for 5 bins in halo mass.
Although according to our classification scheme thresholds can span the 
range $T_j\in[0.2, 1]$, in practice this interval is not always fully probed, 
either because of limited number statistics or because certain threshold values are never
reached by the computed probabilities.

As a general trend and as expected, sensitivity decreases with threshold while 
precision increases, however, there are distinctive features for each class. 
\begin{itemize}
    \item Cluster members: in general at fixed threshold both, sensitivity and precision decrease with
    increasing halo mass. This is a result of a decreasing fraction of cluster galaxies as a function of 
    halo mass in the zone of the PPSD we probe (Fig. \ref{fig:fractions}).    
    \item Backsplash: there is not a clear halo mass dependence as is the case of cluster members, furthermore, for $T>0.4$, the sensitivity has no mass dependence at all, and for the precision differences are very small.
    These trends are consistent with a population that shows, in practice, no dependence with halo mass in 
    terms of both, its location in the PPSD (Fig. \ref{fig:medians}), and its relative abundance (Fig. 
    \ref{fig:fractions}).
    \item Recent infallers: this class shows a halo mass dependence at fixed threshold value opposite to that of the cluster member 
    class: a succesful classification of recent infallers galaxies benefits from the larger number of these galaxies 
    in the PPSD of massive haloes (Fig. \ref{fig:fractions}). 
    \item Infalling: at fixed threshold they are better classified in the PPSD of increasingly more massive systems,
    which, as in the recent infallers case, can be understood in terms of an increasing relative abundance with mass (Fig.
    \ref{fig:fractions}).
    \item Interloopers: the sensitivity curves in Fig. \ref{fig:sp_t} cross at a threshold value of $\sim 0.5$, at lower
    values the sensitivity is slightly higher for lesser massive systems, while an opposite and
    stronger behaviour is seen at larger values of the threshold.  At fixed threshold value, the precision
    increases with mass. These galaxies show almost no trend with halo mass in Figs. \ref{fig:medians} and
    \ref{fig:fractions}. 
\end{itemize}

For each of the five halo mass bins used in Fig. \ref{fig:sp_t}, we compute the value of the threshold that 
minimises the distance of the corresponding pair $($sensitivity, precision$)$ with respect to the ideal 
values $(1,1)$. The resulting values of threshold, sensitivity, and precision are shown in Fig. \ref{fig:tps}
as a function of the halo mass. As can be seen in this figure, the mass dependence of the three quantities
is not strong in any case. We perform linear fits of these threshold values as a function of 
$\log(M_{200}/h^{-1}M_{\odot})$, which are shown in Fig. \ref{fig:tps}, and the corresponding best fitting values
are quoted in Table \ref{tab:thresh}.

\begin{table}
    \centering
    \begin{tabular}{l|rc}
         Class $(j)$ & $a_j$ & $b_j$ \\
         \hline
         Cluster members & $-0.03$ & $0.17$ \\
         Backsplash & $0.01$ & $0.19$ \\
         Recent infallers & $0.0$ & $0.18$ \\
         Infalling & $0.03$ & $0.27$ \\
         Interloopers & $-0.04$ & $0.26$ \\
    \end{tabular}
    \caption{Parametres of the best fitting linear threshold-log halo mass 
    relation that maximizes simultaneously both, precision and sensitivity: $T_\mathrm{max}^{(j)}=a_j\, [\log(M_{200}/h^{-1}M_{\odot})-15]+b_j$. See Fig. \ref{fig:tps}.}
    \label{tab:thresh}
\end{table}

We arrive here at our final and preferred classification scheme out of the \rogerii\ computed probabilities
using a mass dependent threshold: a galaxy is classified as being of the predicted $j-$class if: (i) the maximum value of its $p_i$ $(i=1,\ldots,5)$ corresponds to 
$i=j$, and (ii) $p_j\geq T_\mathrm{max}^{(j)}=a_j\, [\log(M_{200}/h^{-1}M_{\odot})-15]+b_j$ where
$a_j$ and $b_j$ take the values in Table \ref{tab:thresh}.

In Fig.~\ref{fig:distros}, we show the PPSD distributions for different host cluster mass bins, considering the true orbital classes (upper panels) and the corresponding predicted classes (lower panels). First, it can be seen that the host cluster mass introduces only mild variations in the regions of the PPSD occupied by the different orbital classes. Second, the \roger{} code successfully recovers the main features of the PPSD distributions associated with each class, showing a weak dependence on the host cluster mass.

We show in Fig. \ref{fig:hist_rp} the distributions of $R_p/R_{200}$ 
for the predicted classes in the control sample, split into the five bins in
halo mass we use in Figs. \ref{fig:sp_t} and \ref{fig:tps}. We also show in this figure the real distributions.
The distribution of projected distances is reasonably well recovered for the predicted cluster members, recent infallers and interloopers. The predicted BS span a narrower interval in this PPSD coordinate, selecting 
galaxies typically in the range $1\leq R_p/R_{200}\leq 2$. Galaxies with projected distances at 
both tails of the distribution are absent from the predicted sample. These are the zones where 
superposition with cluster members (inner tail) and infalling galaxies (outer tail) results in a set of probabilities
for those galaxies that do not allow a clear distinction among the competing classes.
Finally, predicted infalling are selected preferentially at outer distances than their real counterparts.
This is due to confusion with cluster members at $R_p/R_{200}\lesssim 1$ and with backsplash at 
$1\lesssim R_p/R_{200}\lesssim 2R_{200}$.

Complementarily, in Fig. \ref{fig:hist_dv} we show the corresponding distributions of $|\Delta V|/\sigma$. 
The real distributions are reasonably well matched by the distributions of the predicted
 cluster members, backsplash and infalling galaxies. Predicted recent infallers have narrower and shifted to higher velocities, 
$|\Delta V|/\sigma \gtrsim 1$, distributions where the confusion with cluster members is lesser 
important. 
Predicted interlooper's line of sight 
velocity distributions are biased towards higher velocities where the superposition with 
infalling galaxies is not an issue. 

To further assess the classification scheme, Fig. \ref{fig:conf} displays the resulting confusion matrices, 
with real and predicted classes plotted on the $x$ and $y$ axes, respectively. 
Each row shows the distribution of actual classes within a given predicted class, expressed as a fraction.
The general trends found in \citet{delosRios:2021} are present here. Irrespective of the halo mass,
we observe that the predicted classes are affected by the expected sources of contamination: predicted
cluster members are mostly affected by actual recent infallers, predicted backsplash by actual infalling, predicted recent infallers by actual cluster members, predicted infalling by actual interloopers, and predicted interloopers by actual infalling galaxies. Mild tendencies with halo mass are also present in Fig. \ref{fig:conf}:
the innermost classes, cluster members and recent infallers, are predicted best at the lower mass end, on the contrary, the outermost 
classes, infalling and interloopers, are recovered best at the high mass end. Backsplash are the class that is recovered consistently
at a $\sim 0.41$ value across the whole range of halo mass.

\section{\pyroger{}, the python implementation of \rogerii{}}\label{app:pyroger} 

In this work, we present \rogerii{}\footnote{\url{https://github.com/Martindelosrios/pyROGER}}, the second version of the \roger\ code, which has been fully developed in Python, in contrast to the original version, which was implemented in R. The \rogerii\ package can be readily installed using the standard Python package installer, \texttt{pip}, with the following command:
\begin{equation*}
\texttt{pip install pyROGER}
\end{equation*}
This framework allows you to train a model and saved the corresponding files, so anyone can use the trained model without the need of re-training.

For checking the available models you need to load the \pyroger{} library and list the saved models:

\begin{lstlisting}[language=Python]

# First you need to import the roger library and the roger models...
from pyROGER import roger
from pyROGER import models

# Then you can check the available models...

models.list_saved_models()

\end{lstlisting}

Once you know the exact path to the saved models, you need to load it.

\begin{lstlisting}[language=Python]
# For uploading a model just give the path to the saved .joblib file

models.Roger2.train(path_to_saved_model= ['PATH_TO_MODEL/roger2_KNN.joblib',
                                                  'PATH_TO_MODEL/roger2_RF.joblib'])

# To check that the model is ready to be used you can do:

models.Roger2

# and you should see 'Ready to use' after some comments about the model.

\end{lstlisting}

Finally, you can easily use the saved models and classify your galaxies or estimate their probabilities of belonging to a given class.

\begin{lstlisting}[language=Python]
#Finally, for using the model you just do:

# For predicting the classes:
pred_class = models.Roger2.predict_class(data, nmodel = 0)

# For predicting probabilities:
pred_prob = models.Roger2.predict_prob(data, nmodel = 0)

# Here data should be a np.array with shape (ngals, 3) where the first column is the log10 of the cluster mass in M_{sun}/h units. the second column is the cluster-centric distance normalized to R200, and the third column is the relative velocity normalized to sigma:
#     data[:,0] = np.log10(M_{halo})
#     data[:,1] = R / R200
#     data[:,2] = V / sigma

# On the other hand nmodel just indicates the model that will be used to analyze data. In this example nmodel = 0 is the KNN, nmodel = 1 is the random Forest and nmodel = 2 is the SVM.
\end{lstlisting}

In addition, if you know the real classes of your galaxies, you can compute the corresponding confusion matrix.

\begin{lstlisting}[language=Python]
confusion_matrix,_ = models.Roger2.confusion_matrix(real_class, pred_class)
\end{lstlisting}

This confusion matrix, in turn, can be used for improving the properties of your sample as demonstrated in \citet{Martinez:2025}.

All the pre$-$trained models used in this work can be downloaded from \url{https://zenodo.org/records/21933562}.
With these models you can reproduce all the analysis and plots presented in this work following the python notebook available at \url{https://github.com/Martindelosrios/pyROGER/tree/dev/EXAMPLES}.

It is worth to remark that all models that belongs to ROGERv1 are also available in the \pyroger{} library.

\section{Conclusions}
\label{sec:conclu}

In this work, we present an updated version of the \roger{} code, called \rogerii{}, for performing an orbital classification of galaxies around clusters. 

Compared to the original version, we extend the range of host cluster masses over which the code can be applied, from $10^{15}h^{-1}M_{\odot}$ down to $10^{13.5}h^{-1}M_{\odot}$. In view of the new galaxy surveys expected to become available in the coming years, this extension opens the possibility of performing orbital classifications for thousands of additional galaxies.

 In addition, we incorporate the host cluster mass as a third parameter in the classification. As expected, given the approximate self-similarity of dark matter haloes and the normalization of both axes of the PPSD, the cluster mass introduces only mild modifications to the regions of the PPSD occupied by the different orbital classes. Nevertheless, including the host cluster mass as an additional feature improves the robustness of the classification.

Furthermore, we increase the size of the training set by a factor of three by including the $x$-, $y$-, and $z$-axis projections of each galaxy. This procedure makes the classification more robust to the orientation of the host cluster and reduces its dependence on the particular line of sight adopted.

As described in Section~\ref{sec:model}, the \rogerii{} code provides, for each galaxy, the probability of belonging to each orbital class. Therefore, a final classification requires the definition of a probability threshold. In Section~\ref{sec:test}, we investigate how the sensitivity and precision of the resulting samples vary as a function of this threshold. As expected, we find only a mild dependence of the optimal threshold, and of the corresponding confusion matrices, on the host cluster mass. These results demonstrate the robustness of the method when applied to galaxies residing in clusters spanning a broad range of masses.

It is worth remarking that there are other classification methods similar to \roger{}. In particular, \cite{Farid_2023} developed a machine learning algorithm to classify galaxies in orbiting, infalling and background galaxies. Although they do not use the same classification scheme that us, they provide a comparison with \roger{} showing similar results. Furthermore, during the final stages of development of \rogerii{}, \cite{haggar2026} presented a random forest method for distinguishing backsplash from infalling galaxies in a region between $R_{200}$ and $2\times R_{200}$ around clusters. As expected, if we restrict our method to that region and attempt to classify only those 2 classes, we obtain similar results when using the same set of features. It is worth remarking that, in their work, they show that adding more features may improve the final
results. Nevertheless, we prefer to keep our set of input parameters to those driven mainly by gravitational forces as these will be more robust to the different numerical implementations of cosmological simulations.

We further examine the distributions of the PPSD coordinates for the true and recovered orbital classes in different cluster mass bins. As expected, differences are observed between the true and recovered distributions, primarily due to the intrinsic limitations associated with inferring three-dimensional orbital properties from projected phase-space coordinates. Nevertheless, these discrepancies remain approximately consistent across the different mass bins, providing further evidence for the robustness of the method throughout the explored cluster mass range.

It is worth recalling that a machine learning model such as \rogerii{} is, in the end, 
an optimization problem. Since \rogerii{} was trained on the MDPL2-SAG data, 
any biases or modelling inaccuracies present in the simulation are expected to propagate into 
the final results and affect the predicted probabilities. Nevertheless, the features we used 
for classification were specifically selected for their robustness to differences in the numerical 
implementations adopted in cosmological simulations. In particular, the PPSD coordinates primarily 
depend on the gravitational forces, with only a potentially marginal contribution from additional 
forces arising from hydrodynamical interactions with the gas and stellar components. 
Moreover, the larger volume of MDPL2 compared to hydrodynamical simulations allows us to construct a substantially larger training sample, which can contribute to the robustness of the resulting model.
Among the various factors that may be affecting our model we can mention that $R_{200}$, which is 
used to normalize the PPSD coordinates, and $M_{200}$, may exhibit a mild 
dependence on the halo-finding algorithm used in the different cosmological simulations and in real observations. Furthermore, how the baryonic component was treated by 
SAG has implications at the time of selecting our samples of galaxies. In our case, we use 
central and satellite with stellar masses $M_{\star}\geq 10^9 h^{-1} M_\odot$. Through this 
election the semi-analitic modeling definitely has an impact, however small it may be, 
in the resulting model.
Although we do not expect all these effects to significant affect the final predictions, 
a comprehensive assessment of the application of 
\rogerii{} to different cosmological simulations will be presented in future work.

Finally, we present a Python implementation of the \rogerii{} code, which will be made publicly available. This implementation allows users to classify, in a fast and robust manner, arbitrary galaxy samples using either version of the classifier. It also provides the possibility of training a custom classifier using a different training set, thereby allowing the method to be adapted to alternative datasets or applications.

With the aim of ensuring a fully reproducible analysis, all the plots presented in this paper were generated using a Python Colab notebook, which can be made available upon request.

\section*{Acknowledgements}

This paper has been partially supported with grants from Consejo Nacional de
Investigaciones Cient\'ificas y T\'ecnicas (PIP 11220210100064CO), Argentina,
and Secretar\'ia de Ciencia y Tecnolog\'ia, Universidad Nacional de C\'ordoba (SECYT-UNC, Res. 258/53), Argentina. 
The authors thank Ignacio Germ\'an Alfaro and Yamila Yaryura for their
help with \textsc{MDPL2-SAG} data.  
HJM thanks Tingvall Trio.
The \textsc{CosmoSim} database used in this
paper is a service by the Leibniz-Institute for Astrophysics Potsdam (AIP).  The
\textsc{MultiDark} database was developed in cooperation with the Spanish
MultiDark Consolider Project CSD2009-00064. The authors gratefully acknowledge
the Gauss Centre for Supercomputing e.V. (\url{www.gauss-centre.eu}) and the
Partnership for Advanced Supercomputing in Europe (PRACE, \url{www.prace-ri.eu})
for funding the \textsc{MultiDark} simulation project by providing computing
time on the GCS Supercomputer SuperMUC at Leibniz Supercomputing Centre (LRZ,
\url{www.lrz.de}).

\section*{Data availability}

\rogerii{}~data underlying this article are available in \textsc{Github} at \url{https://github.com/Martindelosrios/pyROGER} and in \textsc{zenodo} at  
\url{https://zenodo.org/records/21933562}. The raw data of the semi-analytic model of galaxy 
formation \textsc{SAG} will be shared on reasonable request to the corresponding author.



\bibliographystyle{mnras}
\bibliography{references}


\bsp	
\label{lastpage}
\end{document}